\documentstyle[aps,prl,epsf,floats]{revtex} 

\draft

\begin{document}
\twocolumn[\hsize\textwidth\columnwidth\hsize\csname
@twocolumnfalse\endcsname 

\title{New Supernova Limit on Large Extra Dimensions}

\author{Steen Hannestad}
\address{NORDITA, Blegdamsvej 17, 2100 Copenhagen, Denmark}

\author{Georg~G.~Raffelt} 
\address{Max-Planck-Institut f\"ur Physik 
(Werner-Heisenberg-Institut), 
F\"ohringer Ring 6, 80805 M\"unchen, Germany}

\date{\today}

\maketitle
                    
\begin{abstract}
  If large extra dimensions exist in nature, supernova (SN) cores will
  emit large fluxes of Kaluza-Klein gravitons, producing a cosmic
  background of these particles with energies and masses up to about
  100~MeV.  Radiative decays then give rise to a diffuse cosmic
  $\gamma$-ray background with $E_\gamma\alt 100~{\rm MeV}$ which is
  well in excess of the observations if more than 0.5--1\% of the SN
  energy is emitted into the new channel.  This argument complements
  and tightens the well-known cooling limit from the observed duration
  of the SN~1987A neutrino burst.  For two extra dimensions we derive
  a conservative bound on their radius of $R \lesssim 0.9 \times
  10^{-4}$~mm, for three extra dimensions it is
  $R \lesssim 1.9 \times 10^{-7}$~mm.
\end{abstract}

\pacs{PACS numbers: 11.10.Kk, 11.25.Mj, 04.50.+h, 97.60.Bw}
\vskip2.0pc]    

%%%%%%%%%%%%%%%%%%%%%%%%%%%%%%%%%%%%%%%%%%%%%%%%%%%%%%%%%%%%%%%%%%%%%%
%% Section I %%%%%%%%%%%%%%%%%%%%%%%%%%%%%%%%%%%%%%%%%%%%%%%%%%%%%%%%%
%%%%%%%%%%%%%%%%%%%%%%%%%%%%%%%%%%%%%%%%%%%%%%%%%%%%%%%%%%%%%%%%%%%%%%

{\it 1.~Introduction.}---The Planck scale of about $10^{19}$~GeV,
relevant for gravitation, is very much larger than the electroweak
scale of about 1~TeV of the particle-physics standard model.  A
radical new approach to solving this notorious hierarchy problem holds
that there could be large extra dimensions, the main idea being that
the standard-model fields are confined to a 3+1 dimensional brane
embedded in a higher dimensional bulk where only gravity is allowed to
propagate~\cite{add98,Antoniadis:1998ig,add99,Han:1999sg,grw99}.  This
concept immediately puts stringent constraints on the size of the
extra dimensions because Newton's law holds at any scale which has
thus far been observed, i.e.\ down to about 1~mm.  Extra dimensions
can only appear at a smaller scale.

For simplicity we follow the usual practice and assume that the $n$
new dimensions form an $n$-torus of the same radius $R$ in each
direction (see, however, Ref.~\cite{kaloper2} for a more general
model).  The Planck scale of the full higher dimensional space,
$M_{{\rm P},n+4}$, can be related to the normal Planck scale,
$M_{{\rm P},4}=1.22\times10^{19}~{\rm GeV}$, by Gauss'
law~\cite{add98}
\begin{equation}
M_{{\rm P},4}^2 = R^n M_{{\rm P},n+4}^{n+2}.
\end{equation}
Therefore, if $R$ is large then $M_{{\rm P},n+4}$ can be much smaller
than $M_{{\rm P},4}$.  If this scenario is to solve the hierarchy
problem then $M_{{\rm P},n+4}$ must be close to the electroweak scale,
i.e.\ $M_{{\rm P},n+4} \lesssim 10$--100~TeV.  This requirement already
excludes $n=1$ because $M_{{\rm P},n+4} \simeq 100$~TeV corresponds to
$R \simeq 10^{8}$~cm.  However, $n \geq 2$ remains possible, and
particularly for $n=2$ there is the intriguing perspective that the
extra dimensions could be accessible to experiments probing gravity at
scales below 1~mm.

Thus far the most restrictive constraint on $M\equiv M_{{\rm P},n+4}$
arises from the observed duration of the supernova (SN)~1987A neutrino
burst~\cite{Cullen:1999hc,SN1987A,Hanhart:2001er,Hanhart:2001fx}.  If
large extra dimensions exist, then the usual 4D graviton is
complemented by a tower of Kaluza-Klein (KK) states, corresponding to
the new available phase space in the bulk.  These KK gravitons would
be emitted from the SN core after collapse, compete with neutrino
cooling, and shorten the observable signal.  This argument has led to
the tight bound $R \lesssim 0.66~\mu$m ($M \gtrsim 31$~TeV) for $n=2$
and $R \lesssim 0.8$~nm ($M \gtrsim 2.75$~TeV) for
$n=3$~\cite{Hanhart:2001fx}.  Cosmological considerations lead to
similar constraints, but the cosmological uncertainties are arguably
larger than those connected with the SN bound.

We presently derive a related SN bound which is independent of the
low-statistics SN~1987A neutrino signal and which is significantly
tighter.  The KK gravitons emitted by all core-collapse SNe over the
age of the universe produce a cosmological background of these
particles.  Later they decay into all standard-model particles which
are kinematically allowed; for the relatively low-mass modes produced
by a SN the only channels are ${\rm KK} \to 2\gamma$, $e^+e^-$ and
$\nu \bar\nu$.  The relevant decay rates are $\tau_{2\gamma} =
\frac{1}{2}\tau_{e^+e^-} = \tau_{\nu \bar\nu} \simeq 6 \times
10^{9}~{\rm yr} \, (m/100~{\rm MeV})^{-3}$ \cite{Han:1999sg}.
Therefore, over the age of the universe a significant fraction of the
produced KK modes has decayed into photons, contributing to the
observed diffuse cosmic $\gamma$-ray background.  We calculate the
present-day contribution to the MeV $\gamma$-ray background for
different numbers and radii of the large extra dimensions.  It turns
out that the measured cosmic $\gamma$-ray background constrains the KK
emission from SNe more tightly than the SN~1987A neutrino signal and
thus provides the most restrictive limit on large extra dimensions.

A note of caution is due here because our arguments rely
on the assumption that KK modes can only decay into standard
model particles on one brane. If there are other branes present
the KK modes can also decay into standard model particles on these,
and the limit is weakened accordingly. 
The argument also relies on the fact that KK modes
cannot decay into other KK modes with smaller bulk momenta. In
models where the extra dimensions are toroidally compactified
this assumption holds because requiring both energy and momentum 
conservation
leads to zero decay probability. In other spaces which are not
translational invariant this argument will in general not be valid.

%%%%%%%%%%%%%%%%%%%%%%%%%%%%%%%%%%%%%%%%%%%%%%%%%%%%%%%%%%%%%%%%%%%%%%
%% Section II %%%%%%%%%%%%%%%%%%%%%%%%%%%%%%%%%%%%%%%%%%%%%%%%%%%%%%%%
%%%%%%%%%%%%%%%%%%%%%%%%%%%%%%%%%%%%%%%%%%%%%%%%%%%%%%%%%%%%%%%%%%%%%%

{\it 2.~Supernova Rate.}---The first ingredient of our calculation is
the SN rate in the universe which is usually expressed in terms of
the ``SN~unit'' or SNu, corresponding to 1~SN per 100~years per
$10^{10}~L_{B\odot}$ with $L_{B\odot}$ the solar luminosity in the
blue spectral band.  Moreover, the average cosmic luminosity density
in the blue is~\cite{Efstathiou} $h\,1.93^{+0.8}_{-0.6} \times 10^8
L_{B\odot}~{\rm Mpc}^{-3}$ where as usual $h$ is the Hubble constant
in units of $100~{\rm km~s^{-1}~Mpc^{-1}}$.  Therefore, in good
approximation 1~SNu corresponds to a spatial density of
$h\,2\times10^{-4}~{\rm yr^{-1}}~{\rm Mpc}^{-3}$.

The observed present-day SN rate depends on the morphological type of
the host galaxy.  Elliptical and S0 galaxies host hardly any
core-collapse SNe while in spiral galaxies of type Sbc--Sd the rate is
$h^2\,(1.8\pm0.6)~{\rm SNu}$ according to~\cite{Turatto} and
$h^2\,3.9~{\rm SNu}$ according to~\cite{Tammann}.  The rate is similar
in Sm, Irr and Pec galaxies, but only about half as large in types
S0a--S0b.  While early-type galaxies (E and S0) are less frequent than
spirals, they are brighter in the blue.  Assuming spirals contribute
at least half the blue luminosity density and in view of the quoted SN
rates we adopt a nominal density of
\begin{equation}\label{eq:SNrate}
R_{\rm SN} = h^3\,2\times10^{-4}~{\rm yr^{-1}}~{\rm Mpc}^{-3}
\end{equation}
for the present-day rate of core-collapse SNe.  This value is almost
identical to the one adopted in~\cite{Woosley} who apparently used
$h=0.5$.

In the past, especially during the first 1~Gyr after galaxy formation,
the SN rate must have been much larger due to the much larger rate of
early star formation. This issue has been broadly discussed in
calculations of the cosmic $\bar\nu_e$ relic flux from SNe that might
be detectable in neutrino observatories such as
Super-Kamiokande. Equation~(\ref{eq:SNrate}) corresponds to an
integrated $\bar\nu_e$ flux at Earth of about $2~\rm cm^{-2}~s^{-1}$,
assuming $h=0.75$, a cosmic age of 12~Gyr, and the emission of
$2\times10^{57}~\bar\nu_e$ per SN.  The larger rates of star formation
in the past increase this flux prediction by factors of up to 100. The
cosmological ``bench-mark case'' of Ref.~\cite{Totani} predicts
$44~\rm cm^{-2}~s^{-1}$ for the integrated $\bar\nu_e$ flux at Earth
while \cite{Kaplinghat} find an upper limit of $54~\rm
cm^{-2}~s^{-1}$, but think that a realistic value is perhaps a factor
of ten smaller.  This lower estimate agrees closely
with~\cite{Hartmann} who find an integrated flux of about $10~\rm
cm^{-2}~s^{-1}$ per neutrino flavor.

In our case of slowly decaying particles, the early SN rate is even
more important because the KK gravitons from early SNe have the
longest time to decay and thus contribute more to the cosmic
$\gamma$-ray background. Therefore, with the constant rate
Eq.~(\ref{eq:SNrate}) we likely underestimate the contribution to the
cosmological $\gamma$-ray background by at least a factor of 10, and
perhaps by as much as a factor of~100.

%%%%%%%%%%%%%%%%%%%%%%%%%%%%%%%%%%%%%%%%%%%%%%%%%%%%%%%%%%%%%%%%%%%%%%
%% Section III %%%%%%%%%%%%%%%%%%%%%%%%%%%%%%%%%%%%%%%%%%%%%%%%%%%%%%%
%%%%%%%%%%%%%%%%%%%%%%%%%%%%%%%%%%%%%%%%%%%%%%%%%%%%%%%%%%%%%%%%%%%%%%

{\it 3.~Emission from SN Core.}---The next ingredient of our
calculation is the emission of KK gravitons from a hot and dense SN
core after collapse.  The main process is nucleonic bremsstrahlung
$N+N\to N+N+{\rm KK}$ which has been discussed in detail in
Ref.~\cite{Hanhart:2001er}. Besides the density, temperature and
chemical composition of the nuclear medium, the energy-loss rate $Q$
into KK states depends on the mass $m$ of these particles.  The mass
of the KK gravitons appears as a consequence of their momentum in the
compactified extra dimensions and as such is a discrete parameter, but
the mode spacings are so small that we may treat $m$ as a continuous
parameter characterizing the tower of KK states that can be produced
by the medium.

On the basis of Ref.~\cite{Hanhart:2001er} we have calculated the
differential emissivity $dQ/dE$ and $dQ/dm$ assuming that the nucleons
are non-relativistic and non-degenerate.  This approximation is only
marginally valid in a hot SN core, but it has no serious bearing on
our results.  A more complete treatment of the nuclear medium is
important for the {\it absolute\/} emission rate, but for the moment
we are only interested in the spectrum which will not significantly
depend on the degree of nucleon degeneracy.  In Fig.~1 we show $dQ/dE$
and $dQ/dm$ for $n=2$ and~3.  Naturally, most KK states are produced
near their kinematic limit; their average speed is about 0.7, i.e.\
typically they are only mildly relativistic. After integrating over
all masses, the average energy of the emitted particles is $\langle E
\rangle = 4.25\,T$ and $ 5.42\,T$ for $n=2$ and 3, respectively, while
conversely, after integrating over all energies, the average mass is
$\langle m \rangle = 2.72\,T$ and $ 3.89\,T$ for $n=2$ and~3.

% --------------------------------------
% Figure 1
% --------------------------------------
\begin{figure}[b]
\begin{center}
\vspace{-0.5truecm}
\epsfysize=10truecm\epsfbox{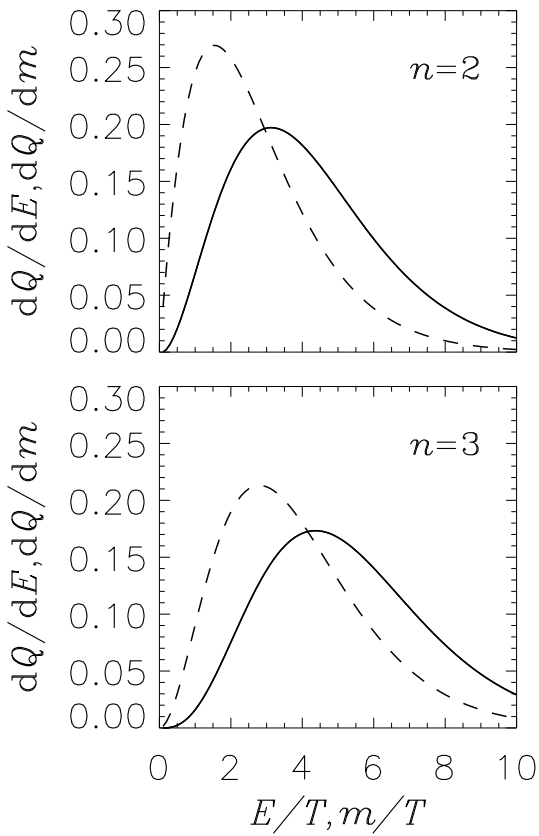}
\vspace{1.5truecm}
\end{center}
\caption{The differential emissivity of KK modes from a
non-degenerate, non-relativistic nuclear medium with a temperature of
30 MeV. The dashed lines are $dQ/dm$ and the full lines are $dQ/dE$ }
\label{fig1}
\end{figure}

The total number of KK gravitons produced in a SN collapse is
$N_{\rm KK} = f_{\rm KK} E_{\rm tot}/\langle E \rangle$, where $E_{\rm
tot}$ is the total SN energy, taken to be $3 \times 10^{53}$~erg, and
$\langle E\rangle$ is the average energy of the emitted KK modes.  The
fraction $f_{\rm KK}$ of the total energy emitted in KK modes is the
main parameter to be constrained by our new argument.

%%%%%%%%%%%%%%%%%%%%%%%%%%%%%%%%%%%%%%%%%%%%%%%%%%%%%%%%%%%%%%%%%%%%%%
%% Background %%%%%%%%%%%%%%%%%%%%%%%%%%%%%%%%%%%%%%%%%%%%%%%%%%%%%%%%
%%%%%%%%%%%%%%%%%%%%%%%%%%%%%%%%%%%%%%%%%%%%%%%%%%%%%%%%%%%%%%%%%%%%%%

{\it 4.~Cosmic Background of Decay Photons.}---Finally we turn
to calculating the present-day photon number density
$n_\gamma$ produced by the decay of KK gravitons from all past SNe.
Formally, this density can be written as
\begin{equation}
\frac{dn_\gamma}{d\epsilon} = 
\int_0^{t_0} dt\, \frac{d^2n_\gamma}{d\epsilon d t},
\end{equation}
where $\epsilon$ is the photon energy and $t_0$ the age of the
universe.  In a standard cosmological model this is
\begin{eqnarray}
\frac{dn_\gamma}{d\epsilon} & = &
\frac{2}{3}H_0^{-1}\int_0^\infty dz\, F(z,\Omega_{\rm M},\Omega_\Lambda)
R_{\rm SN} N_{\rm KK} \nonumber\\
&&\kern2em{}\times
\int_{2 \epsilon}^\infty dm\, \frac{dN_\gamma}{d\epsilon'} 
\frac{dN_{\rm KK}}{dm}\left[1-e^{-t_z/\tau}\right],
\end{eqnarray}
where $\epsilon' = (1+z)\,\epsilon$ with $z$ the cosmic redshift,
$t_z$ is the cosmic time corresponding to redshift $z$, and $\tau$ is
the total KK lifetime which depends on the KK mass $m$.  Notice that
for $\tau \lesssim t_0$ this lifetime disappears from this equation
because essentially all KK modes will have decayed at the present
epoch. The factor $2/3$ arises because the radiative decay rate is
$1/3$ of the total, and because 2~photons are produced per radiative
decay.

Further, $dN_\gamma/d\epsilon$ is the normalized photon energy
distribution from the decay of KK modes, $N_{\rm KK}$ the total number
of KK modes emitted in a given SN, and $dN_{\rm KK}/dm$ their
normalized mass spectrum.  We will always use the limit of
non-relativistic KK particles where $dN_\gamma/d\epsilon = \delta
(\epsilon-m/2)$. This simplification is conservative because we
calculate $N_{\rm KK}$ on the basis of the average emitted energy
$\langle E\rangle$ while the decay photons take each only half of the
mass. On the other hand, we ignore the Lorentz factor for the
decay rate. These two simplifications introduce small errors which 
go in opposite directions.

Finally, $F(z,\Omega_{\rm M},\Omega_\Lambda) = [\Omega_{\rm M} (1+z)^5
+ \Omega_\Lambda (1+z)^2]^{-1/2}$ with $\Omega_{\rm M}$ and
$\Omega_\Lambda$ being the relative contributions of matter and a
cosmological term to the cosmic mass-energy density.  We will use the
standard values $\Omega_{\rm M} = 0.3$, $\Omega_\Lambda = 0.7$ and
$H_0 = 75~{\rm km}~{\rm s}^{-1}~{\rm Mpc}^{-1}$, but our results do
not depend much on the choice of background cosmology.

The present-day diffuse flux from the decaying KK modes obtained in
this way is shown in Fig.~2 for $n=2$ and~3, assuming a SN core
temperature of 30~MeV and $f_{\rm KK} = 1$, i.e.\ that all of the
energy is emitted as KK gravitons.  For comparison we show
a fit to the diffuse $\gamma$-flux observed by the EGRET instrument,
$j(\epsilon) = 2.26 \times 10^{-3}~{\rm cm}^{-2}~{\rm s}^{-1}~{\rm
ster}^{-1}~{\rm MeV}^{-1} (\epsilon/{\rm MeV})^{-2.07}$ 
\cite{egret}.

% --------------------------------------
% Figure 2
% --------------------------------------
\begin{figure}[ht]
\begin{center}
\vspace{-0.5truecm}
\epsfysize=7truecm\epsfbox{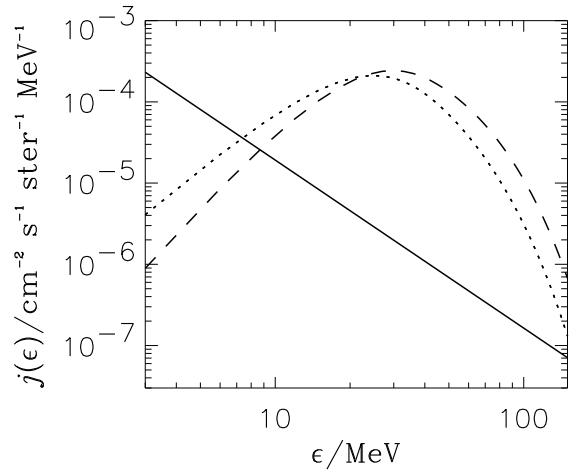}
\vspace{0truecm}
\end{center}
\caption{The present day photon flux due to KK mode decay.  The dotted
line is for $n=2$ 
and the dashed for $n=3$.  The solid line shows the
best fit to the flux observed by EGRET.}
\label{fig2}
\end{figure}

The predicted flux exceeds the observations by about a factor of
100--200 both for $n=2$ and 3.  Therefore, to avoid a conflict with
the data we need to require
\begin{equation}
f_{\rm KK} \lesssim 0.005\hbox{--}0.01.
\end{equation}
This is our main result.  The standard SN~1987A cooling limit amounts
to the requirement $f_{\rm KK} \lesssim 0.5$~\cite{Raffelt:1996wa} so
that our new limit on the energy-loss rate is about two orders of
magnitude more restrictive.

This bound is rather insensitive to the assumed temperature of the
emitting medium. In Fig.~3 we show the maximum allowed value $f_{\rm
max}$ for $f_{\rm KK}$ as a function of $T$. In the entire plausible
range $T \gtrsim 15$~MeV the limit hardly depends on $T$. The reason
is that for a larger temperature the average energy of the emitted KK
states increases, leading to a decrease of their total number. In
addition, the energy of the decay photons is distributed over a
broader range of energies, further decreasing the differential flux.
Therefore, the predicted photon flux is lower, but reaches to larger
energies. On the other hand, the measured $\gamma$-ray flux falls
approximately as $\epsilon^{-2}$, canceling the previous effects.
When $T$ and thus the typical KK masses become too small, the KK
lifetime begins to exceed the age of the universe, decreasing the
predicted flux of decay photons.

% --------------------------------------
% Figure 3
% --------------------------------------
\begin{figure}
\begin{center}
\vspace{-0.5truecm}
\epsfysize=7truecm\epsfbox{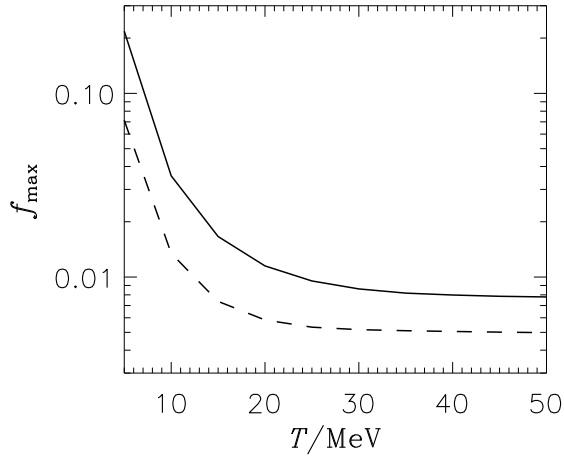}
\vspace{0truecm}
\end{center}
\caption{Maximum allowed value for $f_{\rm KK}$ as a function of the
SN core temperature.  The dotted line is for $n=2$ and the dashed is
for $n=3$.}
\label{fig3}
\end{figure}

%%%%%%%%%%%%%%%%%%%%%%%%%%%%%%%%%%%%%%%%%%%%%%%%%%%%%%%%%%%%%%%%%%%%%%
%% Conclusion %%%%%%%%%%%%%%%%%%%%%%%%%%%%%%%%%%%%%%%%%%%%%%%%%%%%%%%%
%%%%%%%%%%%%%%%%%%%%%%%%%%%%%%%%%%%%%%%%%%%%%%%%%%%%%%%%%%%%%%%%%%%%%%

{\it 5.~Summary.}---We have shown that the decay photons from KK
gravitons emitted by all past SNe exceeds the measured diffuse
$\gamma$-flux unless the fraction of the SN energy released in KK
modes is less than about 0.5--1\% of the total. This limit is very
conservative because we have used the present-day SN rate as
representative for the entire cosmic evolution. A more realistic
assumption could lead to a photon flux of up to a factor of 10--100
larger than our estimate.  Our limit on the fractional SN energy loss
is at least two orders of magnitude more restrictive than the usual
SN~1987A cooling limit. Of course, this latter argument applies to any
invisible energy-loss channel while our new result depends on the
relatively fast radiative decay of the KK modes.

Our new argument implies that if the number of extra dimensions $n$ is
2 or 3, their radius $R$ must be about a factor of 10 smaller than
implied by the SN~1987A cooling limit.  For $n=2$ the SN~1987A limit
is $R \leq 0.71 \times 10^{-3}$~mm \cite{Hanhart:2001er}, so that our
limit becomes roughly $R \lesssim 0.9 \times 10^{-4}$~mm. This
translates into a lower bound on the energy scale of $M \geq 84$~TeV
for $n=2$.  For $n=3$ the cooling limit is $R \leq 0.85 \times
10^{-6}$~mm \cite{Hanhart:2001er} while our new limit is $R \lesssim
0.19 \times 10^{-6}$~mm. The energy scale is then bounded by $M \geq
7$~TeV.

Our new bounds appear to be the most restrictive limits on large extra
dimensions, except for limits which invoke potentially more uncertain
early-universe arguments. Our argument does not depend on the low
counting statistics of the SN~1987A neutrino signal, but shares all
uncertainties related to calculating the KK graviton emission from a
dense nuclear medium.  If the relic neutrinos from all past SNe were
to be observed in a neutrino observatory, such a measurement would pin
down the cosmic SN rate and would allow us to improve our limit
accordingly.

Finally, we stress that the limit we have derived 
(as well as all other astrophysical and cosmological limits)
can be avoided
in some models where compactification of the extra dimensions is
non-toroidal \cite{kaloper2}, or for example in the model
of Dvali et al. \cite{Dvali:2001gm}, 
where a Ricci term on the brane suppresses the
graviton emission rates.

%%%%%%%%%%%%%%%%%%%%%%%%%%%%%%%%%%%%%%%%%%%%%%%%%%%%%%%%%%%%%%%%%%%%%%
%% Acknowledgments %%%%%%%%%%%%%%%%%%%%%%%%%%%%%%%%%%%%%%%%%%%%%%%%%%%
%%%%%%%%%%%%%%%%%%%%%%%%%%%%%%%%%%%%%%%%%%%%%%%%%%%%%%%%%%%%%%%%%%%%%%
 
{\it Acknowledgments.}---In Munich, 
this work was partly supported by the Deut\-sche
For\-schungs\-ge\-mein\-schaft under grant No.\ SFB 375. 
We thank Shmuel Nussinov, Gia Dvali and Sacha Davidson
for interesting discussions
on the possibility of bulk decays of KK modes.

%%%%%%%%%%%%%%%%%%%%%%%%%%%%%%%%%%%%%%%%%%%%%%%%%%%%%%%%%%%%%%%%%%%%%%
%% References %%%%%%%%%%%%%%%%%%%%%%%%%%%%%%%%%%%%%%%%%%%%%%%%%%%%%%%%
%%%%%%%%%%%%%%%%%%%%%%%%%%%%%%%%%%%%%%%%%%%%%%%%%%%%%%%%%%%%%%%%%%%%%%

\end{document}